\documentclass[conference]{IEEEtran}
\usepackage{tikz}
\usepackage{float}
\usepackage{mcite}
\usepackage{graphicx}
\usepackage{amsmath}
\usepackage{amssymb}
\usepackage{bm}
\usepackage{upgreek}
\usepackage[colorlinks=true, allcolors=blue]{hyperref}
\usepackage{subfigure}
\usepackage{array}
\usepackage{threeparttable}
\usepackage{booktabs}

\definecolor{linkcolor}{RGB}{92,92,192}
\definecolor{blackcolor}{RGB}{0,0,0}
\hypersetup{colorlinks=true,linkcolor=blackcolor,citecolor=blackcolor,urlcolor=linkcolor}

\newcommand{\tref}[1]{Tab.~\ref{#1}}
\newcommand{\fref}[1]{Fig.~\ref{#1}}

\IEEEoverridecommandlockouts

\begin{document}

\title{Scalable Non-Cartesian Magnetic Resonance Imaging with R2D2}

\author{\IEEEauthorblockN{
Yiwei Chen$^1$, Chao Tang$^{1,2}$,
Amir Aghabiglou$^1$, Chung San Chu$^1$, 
Yves Wiaux$^1$\IEEEauthorrefmark{2}\thanks{The work was supported by EPSRC under grants EP/T028270/1 and ST/W000970/1. Computing \mbox{resources} came from the Cirrus UK National Tier-2 HPC Service at EPCC (http://www.cirrus.ac.uk) funded by the University of Edinburgh and EPSRC (EP/P020267/1). 
}}
\IEEEauthorblockA{$^1$Institute of Sensors, Signals and Systems, Heriot-Watt University, Edinburgh EH14 4AS, United Kingdom 
\\$^2$EPCC, University of Edinburgh, Edinburgh EH8 9BT, United Kingdom
}
\IEEEauthorblockA{Email: \IEEEauthorrefmark{2}y.wiaux@hw.ac.uk}
}


%

\maketitle

\begin{abstract}

We propose a new approach for non-Cartesian magnetic resonance image reconstruction. While unrolled architectures provide robustness via data-consistency layers, embedding measurement operators in Deep Neural Network (DNN) can become impractical at large scale. Alternative Plug-and-Play (PnP) approaches, where the denoising DNNs are blind to the measurement setting, are not affected by this limitation and have also proven effective, but their highly iterative nature also affects scalability. To address this scalability challenge, we leverage the ``Residual-to-Residual DNN series for high-Dynamic range imaging (R2D2)'' approach recently introduced in astronomical imaging. R2D2's reconstruction is formed as a series of residual images, iteratively estimated as outputs of DNNs taking the previous iteration's image estimate and associated data residual as inputs. The method can be interpreted as a learned version of the Matching Pursuit algorithm. We demonstrate R2D2 in simulation, considering radial k-space sampling acquisition sequences. Our preliminary results suggest that R2D2 achieves: (i) suboptimal performance compared to its unrolled incarnation R2D2-Net, which is however non-scalable due to the necessary embedding of NUFFT-based data-consistency layers; (ii) superior reconstruction quality to a scalable version of R2D2-Net embedding an FFT-based approximation for data consistency; (iii) superior reconstruction quality to PnP, while only requiring few iterations.

\end{abstract}

%
\IEEEpeerreviewmaketitle

\section{Introduction}
Magnetic Resonance Imaging (MRI) enables high-precision reconstruction of structures and organs within the human body via the interaction of magnetic fields and radio waves. In MRI, signals are measured in the spatial frequency domain, namely k-space, using multiple receiver coils. The complex-valued MR image is reconstructed based on the inverse Fourier transform. For this proof of concept and without loss of generality, similarly to \cite{hyun2018deep}, we consider real-valued positive images, which would correspond to the magnitude of complex-valued MR images. We also restrict the study to a single-coil setting. We consider non-Cartesian k-space sampling, more precisely radial sampling sequences, which are commonly used and possess advantages such as reduced motion artifacts in dynamic scenes \cite{wright2014non}. 
The acquisition process can thus be formulated as
\begin{equation}
\bm{y}=\bm{\mathsf{\Phi}} \bm{\bar{x}} +\bm{n},
\label{eq:forward}
\end{equation}
where $\bm{y} \in \mathbb{C}^M$ denotes the k-space measurements, $\bm{\mathsf{\Phi}}\colon \mathbb{R}^N \to \mathbb{C}^M$ is a non-uniform Fourier sampling measurement operator, $\bm{\bar{x}} \in \mathbb{R}^{N}_{+}$ is the Ground Truth (GT) image and $\bm{n} \in \mathbb{C}^M$ represents a complex-valued Gaussian random noise following $\mathcal{N}(0,\tau^2)$. 
$\bm{\mathsf{\Phi}}$ can be implemented via the Non-Uniform Fast Fourier Transform (NUFFT), \emph{i.e.}~as the product of $\bm{\mathsf{U,F,Z}}$, where $\bm{\mathsf{Z}}$ is a zero-padding operator, $\bm{\mathsf{F}}$ denotes the Fast Fourier Transform (FFT), 
and $\bm{\mathsf{U}}$ is an interpolation operator. 

To accelerate the acquisition process, reducing the number of k-space measurements is common. DNN models like PnP algorithms \cite{kamilov2023plug} and unrolled networks \cite{ramzi2022nc} have gained attention for their promising performance on undersampled MRI data. PnP algorithms, bridging optimization theory and deep learning, train denoising DNNs deployed within optimization algorithms to replace handcrafted regularization operators. While effective and unaffected by measurement settings, they face scalability challenges due to their highly iterative nature. Unrolled DNNs ensure the consistency between the reconstructions and measurements by mirroring optimization algorithm iterations across layers, achieving high imaging precision in non-Cartesian MRI. However, in large-dimensional settings, typically when a high number of acquisition coils are involved or in 3D or 4D dynamic MRI \cite{stankovic20144d}, embedding NUFFT operators into DNN architectures entails large computational cost and memory requirements at both training and inference stages. These can rapidly become prohibitive due to corresponding limitation of GPU hardware \cite{ramzi2022nc}.

Recently, the R2D2 approach, a learned variant of the Matching Pursuit approach \cite{mallat1993matching}, 
has been demonstrated to deliver a new regime of joint image reconstruction quality and speed in radio astronomical imaging \cite{aghabiglou2023deep, aghabiglou2023ultra, aghabiglou2024R2D2}. R2D2's reconstruction is formed as a series of residual
images, iteratively estimated as outputs of DNNs taking the previous iteration’s image estimate and associated back-projected data residual as inputs. In contrast with an unrolled DNN approach, R2D2 thus utilizes a series of DNNs 
without embedding the measurement operator into their individual architectures. Interestingly, the number of required networks (iterations) was shown to be extremely small compared to the typical number of iterations of a PnP algorithm.

We propose two unrolled variants of the R2D2 model, named R2D2-Net (FFT) and R2D2-Net (NUFFT), embedding distinct data-consistency layers, respectively based on the NUFFT  and an FFT approximation. 
Preliminary results suggest that R2D2 naturally exhibits suboptimal performance compared to the R2D2-Net (NUFFT), which, however, is non-scalable due to the essential embedding of NUFFT-based data-consistency layers. However, R2D2 largely outperforms the scalable R2D2-Net (FFT) variant. With respect to the state of the art, R2D2 also largely outperforms the (non-scalable) unrolled network named NC-PDNet, also embedding the NUFFT \cite{ramzi2022nc}, and a (non-scalable) PnP benchmark \cite{Matthieu2021}.


\section{Methodology}\label{sec:method}
\subsection{R2D2 approach}
Given an image estimate $\bm{x}$, the back-projected data residual denoted by $\bm{r}$, also named the
data-consistency term \cite{sriram2020end}, is given by
\begin{equation}
\bm{r}=\bm{x}_{\text{d}}-\kappa \operatorname{Re}\{\bm{\mathsf{\Phi}^{\dagger}} \bm{\mathsf{\Phi}} \boldsymbol{x}\},
\label{Eq:redisual_cal}
\end{equation}
where $\bm{x}_{\text{d}}=\kappa \operatorname{Re} \{ \bm{\mathsf{\Phi}^{\dagger}}\bm{y}\}$ is the data back-projected from the raw measurements. Here, $\kappa=1 / \max (\operatorname{Re}\{\bm{\mathsf{\Phi}^{\dagger}} \bm{\mathsf{\Phi}}\boldsymbol{\delta}\} )$ is the normalization parameter with $\boldsymbol{\delta}$ being the Dirac image with the central pixel value as 1 and others as 0. 

The R2D2 approach aims to utilize the previous image estimate $\boldsymbol{x}^{i-1}$ and corresponding $\bm{r}^{i-1}$ to reduce the discrepancies between the image estimate and GT image by a series of DNNs, denoted as $\{\bm{\mathsf{G}}_{\bm{\theta_i}}\}_{1\leq i \leq I}$, with learnable parameters denoted as $\{\bm{\theta_i}\}_{1\leq i \leq I}$. The iteration structure reads
\begin{equation}
\boldsymbol{x}^{i}=\boldsymbol{x}^{i-1} + \bm{\mathsf{G}}_{\bm{\theta_i}}(\bm{r}^{i-1} , \boldsymbol{x}^{i-1}),
\label{Eq:res}
\end{equation}
with the initialized image estimate and back-projected data residual as $\boldsymbol{x}^{0}=\boldsymbol{0}$ and $\boldsymbol{r}^{0}=\bm{x}_{\text{d}}$, respectively. DNNs are trained sequentially. Specifically, for the $i$-th DNN, the loss function is defined as:
\begin{equation}
\underset{\bm{\theta}_i}{\min} \frac{1}{K}\sum_{k=1}^{K}\lVert \bm{\bar{x}}_{k} -[\boldsymbol{x}^{i-1}_{k} + \bm{\mathsf{G}}_{\bm{\theta_i}}(\bm{r}^{i-1}_{k} , \boldsymbol{x}^{i-1}_{k})]_{+} \rVert_{1},
\label{Eq:loss}
\end{equation}
where $\lVert \cdot \rVert_{1}$ is the L1 norm, $K$ is the number of training samples, and $[\cdot]_{+}$ denotes the projection onto the positive orthant $\mathbb{R}^{N}_{+}$. 

At the inference stage, the last output of the DNN series denoted as $\boldsymbol{x}^{I}$ gives the final reconstruction. It is noteworthy that the first iteration of R2D2 is equivalent to a standard end-to-end learning approach.

\subsection{DNN structure}
The standard U-Net \cite{ronneberger2015u} serves as the foundational structure for DNNs, featuring both a contracting and expanding path. The contracting path integrates multiple convolutional and pooling layers to gradually reduce the spatial dimensions of the input image. Conversely, the expanding path incorporates convolutional and upsampling layers to progressively upsample the feature maps, resulting in an output with identical dimensions to the input image. Skip connections are added to link layers between the contracting and expanding paths.

\subsection{R2D2-Net}
Formally, an unrolled variant of R2D2, named R2D2-Net \cite{aghabiglou2023ultra}, can be developed by unrolling the R2D2 approach itself, with a predetermined number of internal iterations. Two implementations, distinguished by data-consistency computation, are considered.
The first, denoted by R2D2-Net (NUFFT), uses the exact measurement operator based on the NUFFT to calculate the data consistency. To promote scalability, we employ an FFT approximation of the exact measurement operator to estimate the data consistency. Firstly, the point spread function image is calculated as $\bm{\mathbf{h}}=\kappa \operatorname{Re}\{\bm{\mathsf{\Phi}^{\dagger}} \bm{\mathsf{\Phi}} \boldsymbol{\delta}\} \in \mathbb{R}^N$. Secondly, we replace $\kappa \operatorname{Re}\{\bm{\mathsf{\Phi}^{\dagger}} \bm{\mathsf{\Phi}}\}$ by $\operatorname{Re}\{\bm{\mathsf{F}^{\dagger}}(\bm{\mathsf{F}}\bm{\mathbf{h}})\bm{\mathsf{F}}\}$, where the FFT operator and its adjoint are denoted by $\bm{\mathsf{F}}$ and $\bm{\mathsf{F}^{\dagger}}$, respectively. Note that $\bm{\mathsf{F}}\bm{\mathbf{h}}$ is precomputed once and stored. Although this mapping to the back-projected data space is approximate, it is fast and memory-efficient. This substitution results in a scalable unrolled DNN denoted by R2D2-Net (FFT).

\subsection{Normalization procedures}
Normalization procedures are employed to avoid generalizability issues arising from large variations in pixel value ranges and stabilize the training process. An iteration-specific normalization factor for the $i$-th network is denoted as $\alpha^{i-1}$ and given by the mean pixel value of the previous estimate $\boldsymbol{x}^{i-1}$. Note that $\alpha^{0}$ is the mean pixel value of $\bm{x}_{\text{d}}$. In training,
$\bm{\bar{x}}$, $\bm{x}^{i-1}$ and $\bm{r}^{i-1}$ are divided by $\alpha^{i-1}$. At the inference stage, the normalization mapping: $\bm{\mathsf{G}}_{\bm{\theta_i}}(\cdot) \mapsto \alpha^{i-1}\bm{\mathsf{G}}_{\bm{\theta_i}}(\cdot/\alpha^{i-1})$ is applied to each subnetwork to make the input normalized and the output denormalized accordingly.

\subsection{Related works}
Firstly, compared to PnP algorithms, DNNs in R2D2 aim to learn high-level features rather than being trained solely as denoisers \cite{kamilov2023plug}.
Secondly, in contrast to unrolled DNNs \cite{sriram2020end,ramzi2022nc}, R2D2 externalizes the data consistency calculation from the network structure. This strategy relieves the significant computational cost of NUFFT in training, leading to better scalability. Formally, R2D2-Net shares the same network architecture as NC-PDNet \cite{ramzi2022nc}, with the key distinction being that the core subnetwork architecture of R2D2-Net is a U-Net instead of a shallow convolution neural network. Furthermore, R2D2-Net introduces a novel variant employing the FFT approximation to enhance scalability.

\section{Data Simulation}\label{sec:simulation}

\subsection{Denoised GT images}
\begin{figure}
  \centering
\includegraphics[width=0.95\linewidth]{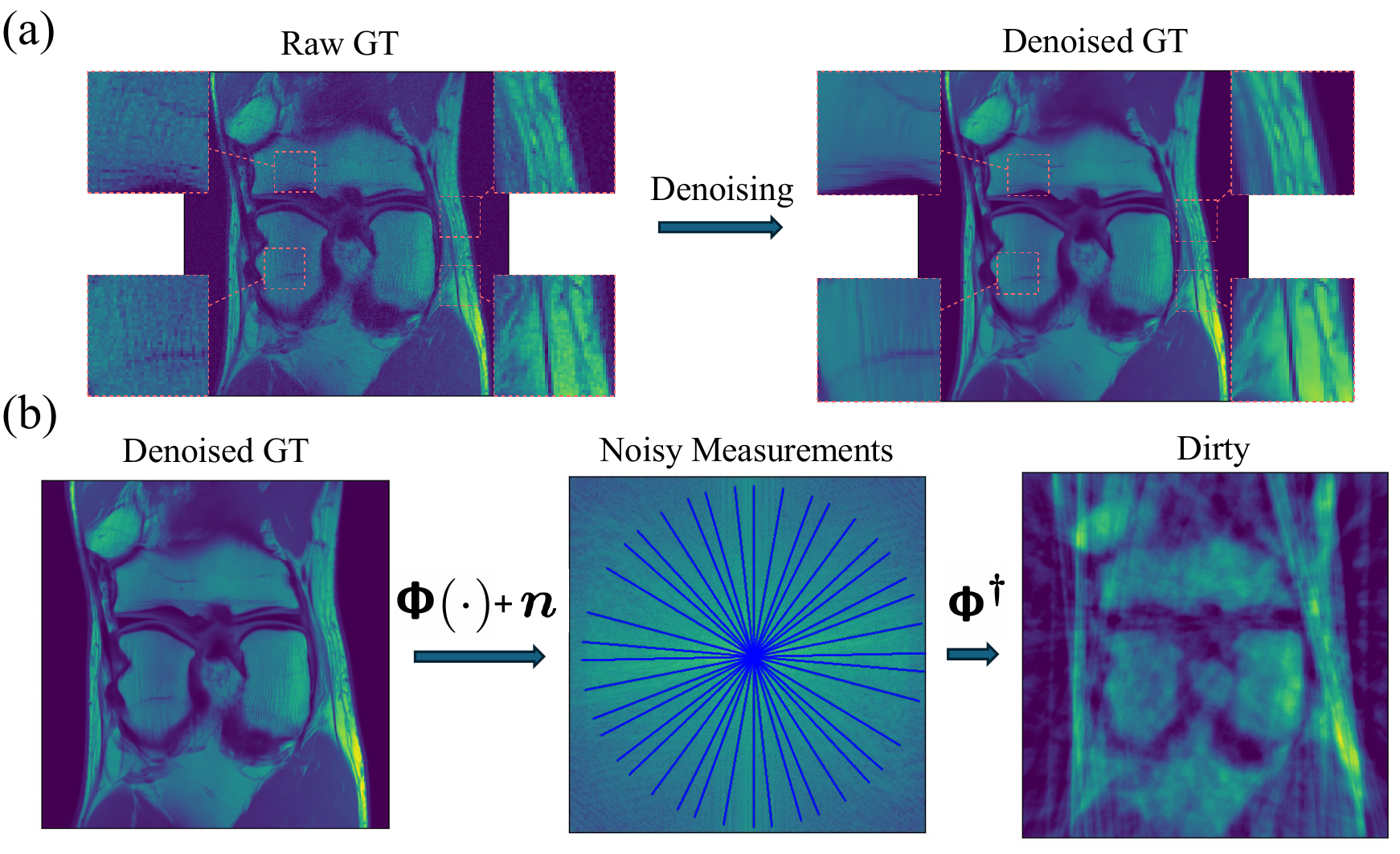}
  \vspace{-10pt}
  \caption{\textbf{Data simulation process.} (a) Denoising the raw GT image; (b) Generating the simulated measurements and back-projected image.}
  \label{fig:dataset}
  \vspace{-10pt}
\end{figure}

We adopt the magnitude MR images with a size of $320 \times 320$ from the FastMRI single-coil knee dataset \cite{zbontar2018fastMRI}. As depicted in \fref{fig:dataset} (a), the raw GT image exhibits noticeable noise. To improve the image quality and generate k-space measurements with diverse noise levels, the raw images undergo preprocessing by the denoising network SCUNet \cite{zhang2023practical}, followed by soft-thresholding to remove residual backgrounds. The denoised image is then normalized with intensities ranging from 0 to 1. We split denoised images into training and validation datasets consisting of 25743 and 6605 images, respectively.

\subsection{Radial sampling}
Radial sampling starts at the centre of k-space and extends outward in an angular fashion. The radial sampling trajectory in k-space is formulated as
$k_x(n)=r  \cos \left(\alpha_n\right)$, $k_y(n)=r  \sin \left(\alpha_n\right)$, 
where $k_x(n)$ and $k_y(n)$ are the Cartesian coordinates of the sampled point at the $n$-th spoke and $r$ represents the radial distance from the k-space origin with $r \in \mathbb{Z}, -R \leq r \leq R$. Generally, for a square image, $R$ is the unilateral dimension of the image \cite{ramzi2022nc}. $\alpha_n$ is the angle of the $n$-th spoke with respect to the $k_x$ axis. The golden angle, $\alpha_g = 111.25^\circ$ for the 2D case \cite{wright2014non}, is a common choice for the specific angular increment to achieve an efficient and relatively uniform angular distribution for the radial sampling trajectory, resulting in $\alpha_n = n \alpha_g$. 

\subsection{Density compensation}
During the measurement process of radial sampling, large intensities are allocated to the densely sampled region at the centre of the k-space, which leads to a back-projected image with abnormally large values. Density Compensation (DC), as proposed in \cite{pipe1999sampling}, calculates factors in k-space that evenly weigh different sample locations by iteratively applying the interpolation matrix and the adjoint interpolation matrix over $m$ iterations. In our implementation, $m$ is set to be 10 following \cite{ramzi2022nc}. After obtaining the DC weights denoted by $\boldsymbol{d}$, we multiply the k-space data by $\boldsymbol{d}$ before the back-projection.

\subsection{Acceleration factor}
In data acquisition, oversampling can be applied along each spoke without increasing the scan time. The number of spokes, denoted by $N_{\text{s}}\in \mathbb{Z}_{+}$, determines the sparsity of the k-space sampling, directly related to the speed of data acquisition, but with sparser sampling implying a more challenging inverse problem for image reconstruction. 
Following \cite{ramzi2022nc}, we define an Acceleration Factor (AF) for 2D radial sampling based on $N_{\text{s}}$ as ${\rm AF} = \sqrt{N}/N_{\text{s}}$. In previous studies, AF was set to be a constant, \emph{e.g.}~4 or 8. Aiming to develop a model adaptable for a range of AFs, and to demonstrate R2D2's performance at high AFs, the number of spoke is left variable during training. 

\subsection{Noise}\label{S:noise}
In our dataset, the highest intensity of GT images is 1 due to the normalization. The faintest image intensity, denoted by $\sigma$, is the standard deviation of the background Gaussian noise in the image domain, which is predicated on the assumption that all predictable intensities surpass the noise level. The Dynamic Range (DR) is defined as the ratio between the intensities of the maximum and faintest features, which in our dataset is specifically the reciprocal of $\sigma$. We configured the DR to span from $10$ to $10^4$, covering both low- and high-noise samples, facilitating robust evaluation of imaging methods. 
Following \cite{wilber2023scalable}, we model the relationship between the additive noise in the k-space and its back-projection in the image domain as $\tau = \sqrt{2{\rm L^2/L_{p}}}\sigma_n$,
where $\rm L$ and $\rm L_{\text{p}}$ are the spectral norm of the measurement operator when the weighting is applied once and twice, respectively. Setting $\sigma_n=\sigma$ ensures that the Gaussian noise in the synthetic measurements preserves the DR of the GT images.

\begin{figure*}[ht] 
\centering
\begin{tabular}{m{2.1cm}<{\centering} m{2.1cm}<{\centering} m{2.1cm}<{\centering} m{2.1cm}<{\centering} m{2.1cm}<{\centering} m{2.1cm}<{\centering}m{2.1cm}<{\centering}}
 \includegraphics[width=0.14\textwidth]{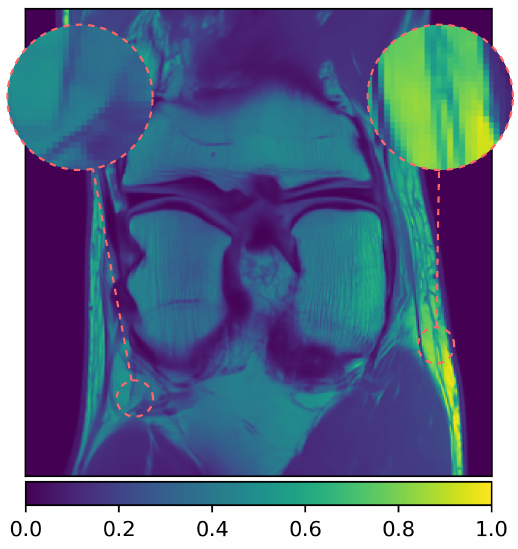} &
 \includegraphics[width=0.14\textwidth]{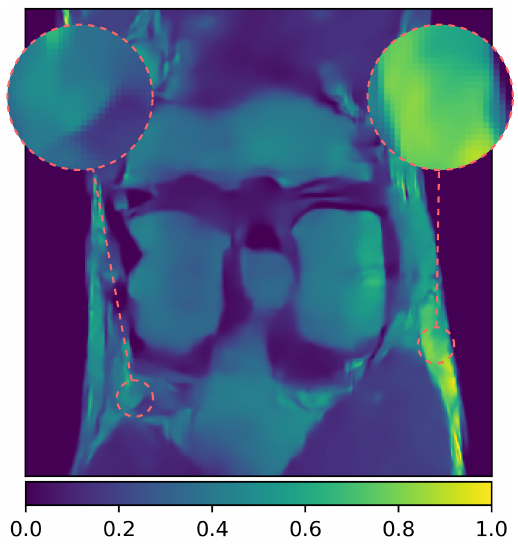}&
\includegraphics[width=0.14\textwidth]{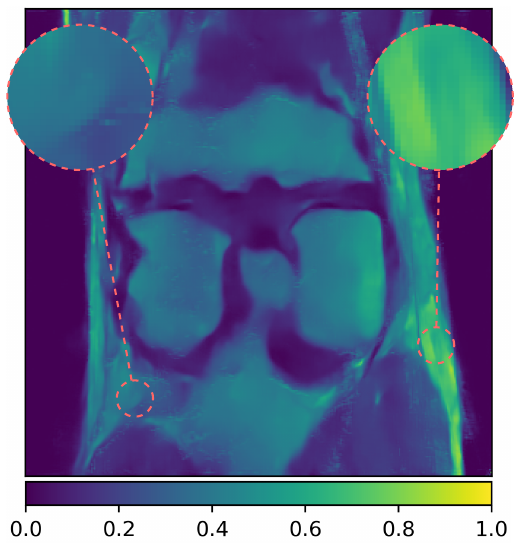}&
\includegraphics[width=0.14\textwidth]{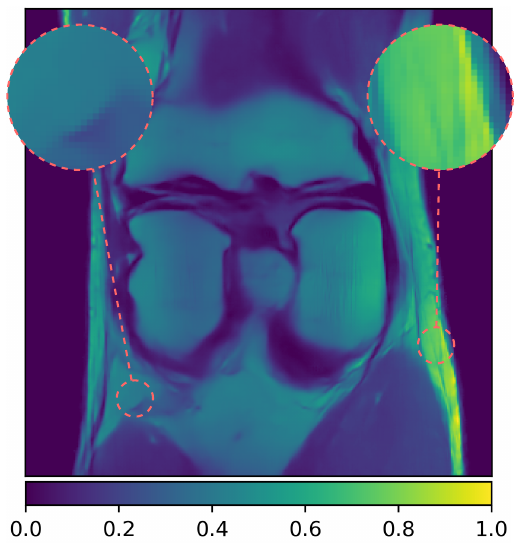}&
\includegraphics[width=0.14\textwidth]{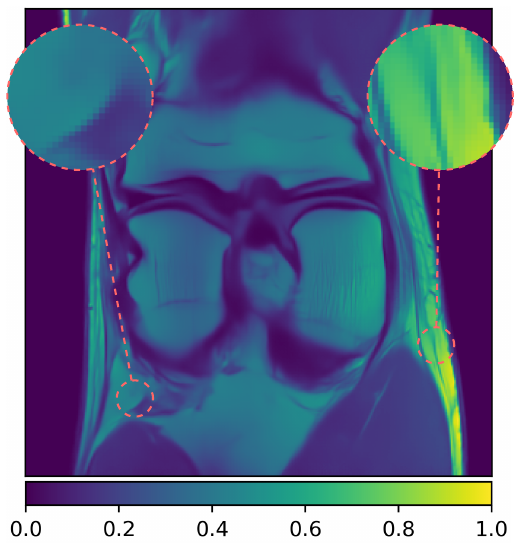}&
\includegraphics[width=0.14\textwidth]{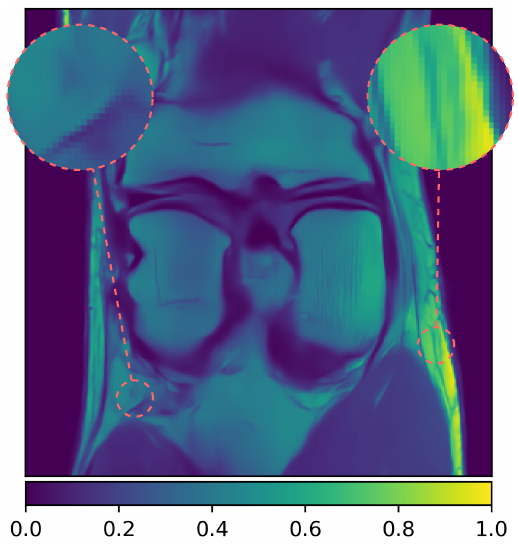} &
\includegraphics[width=0.14\textwidth]{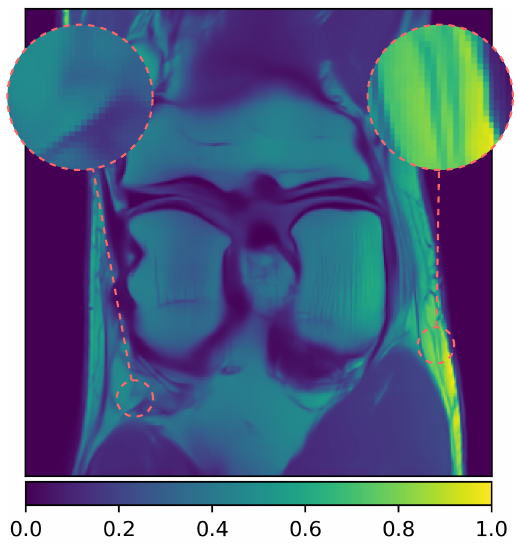} 
 \\
\includegraphics[width=0.137\textwidth]{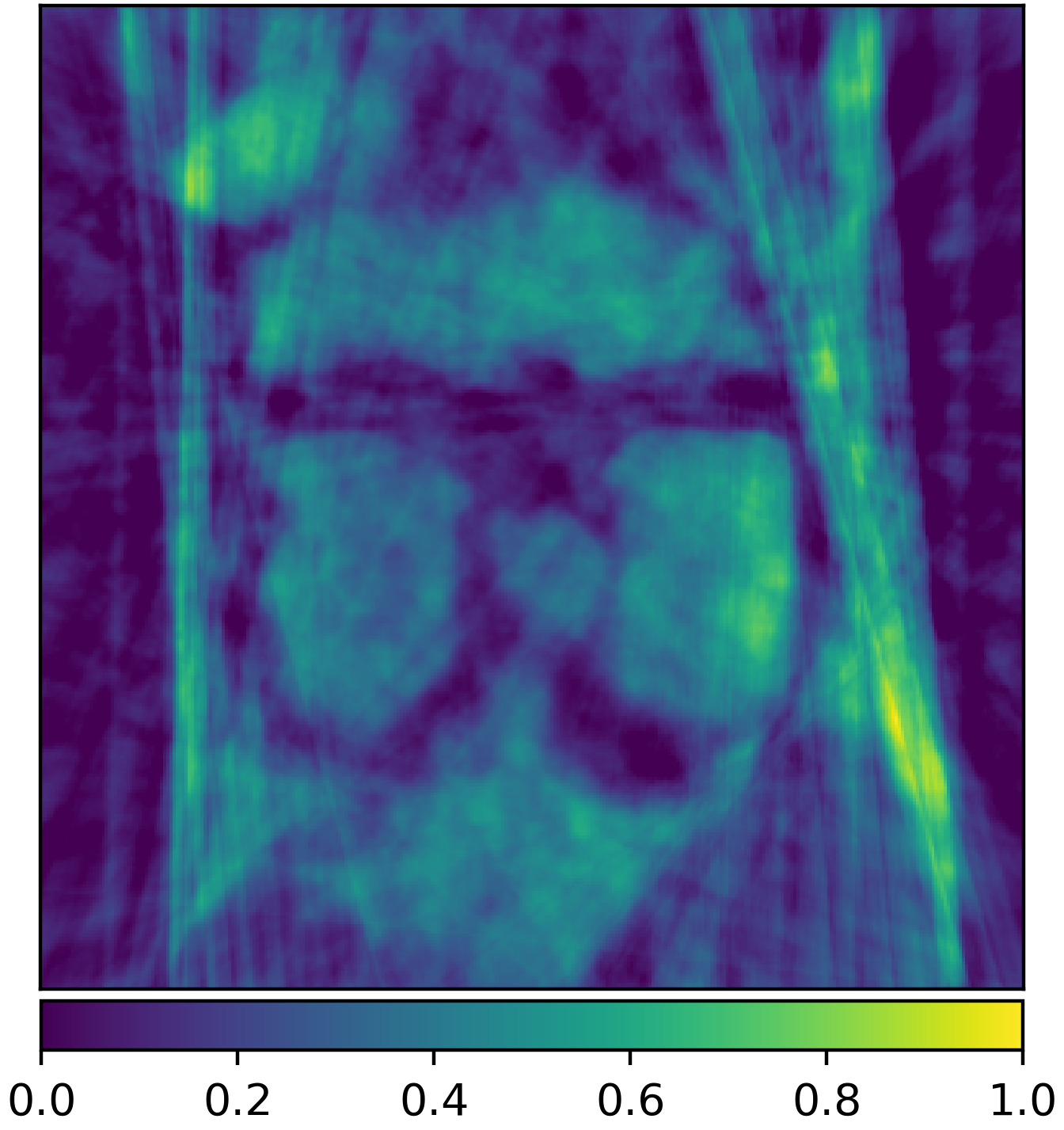} &
\includegraphics[width=0.14\textwidth]{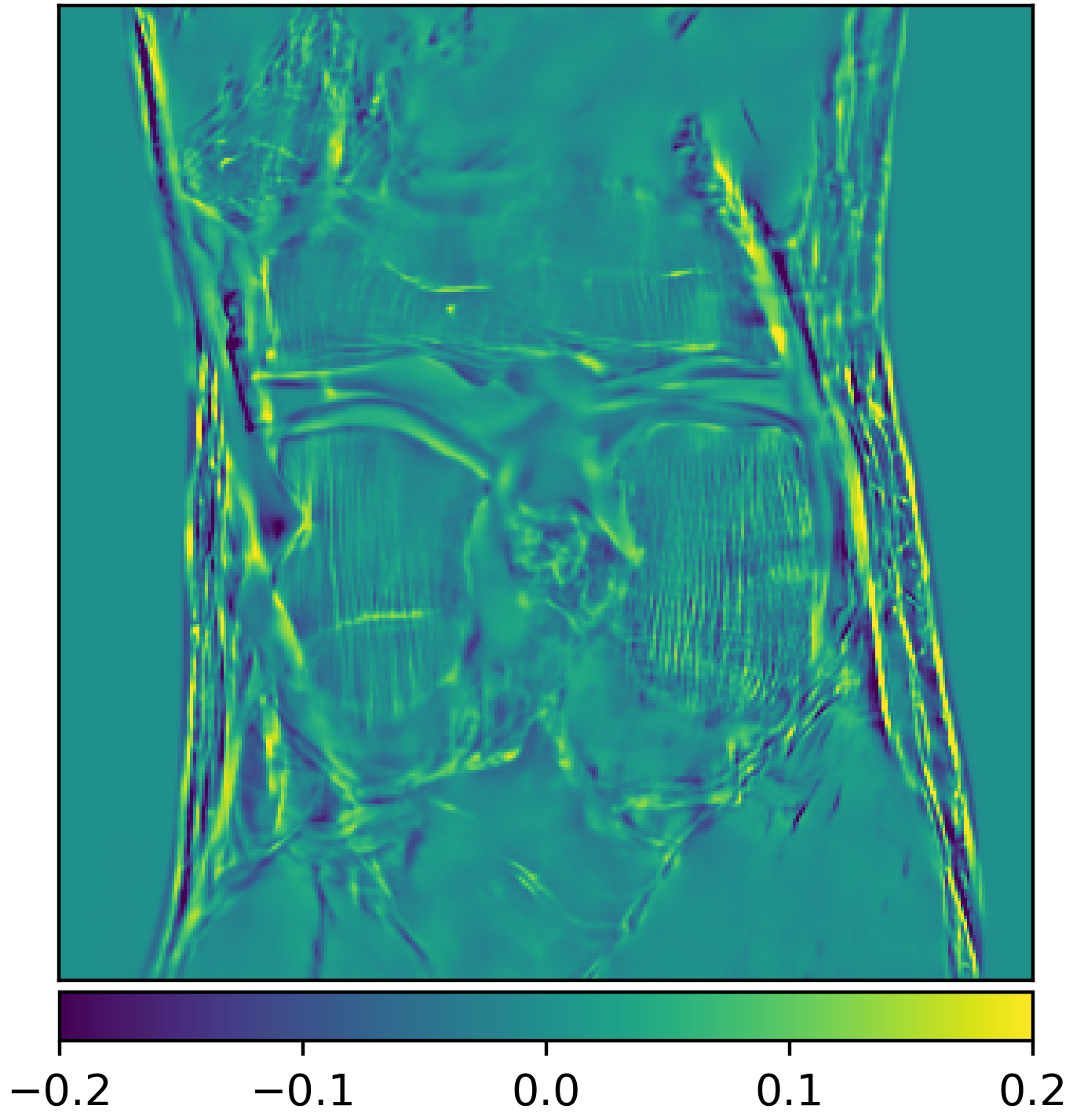}&
\includegraphics[width=0.14\textwidth]{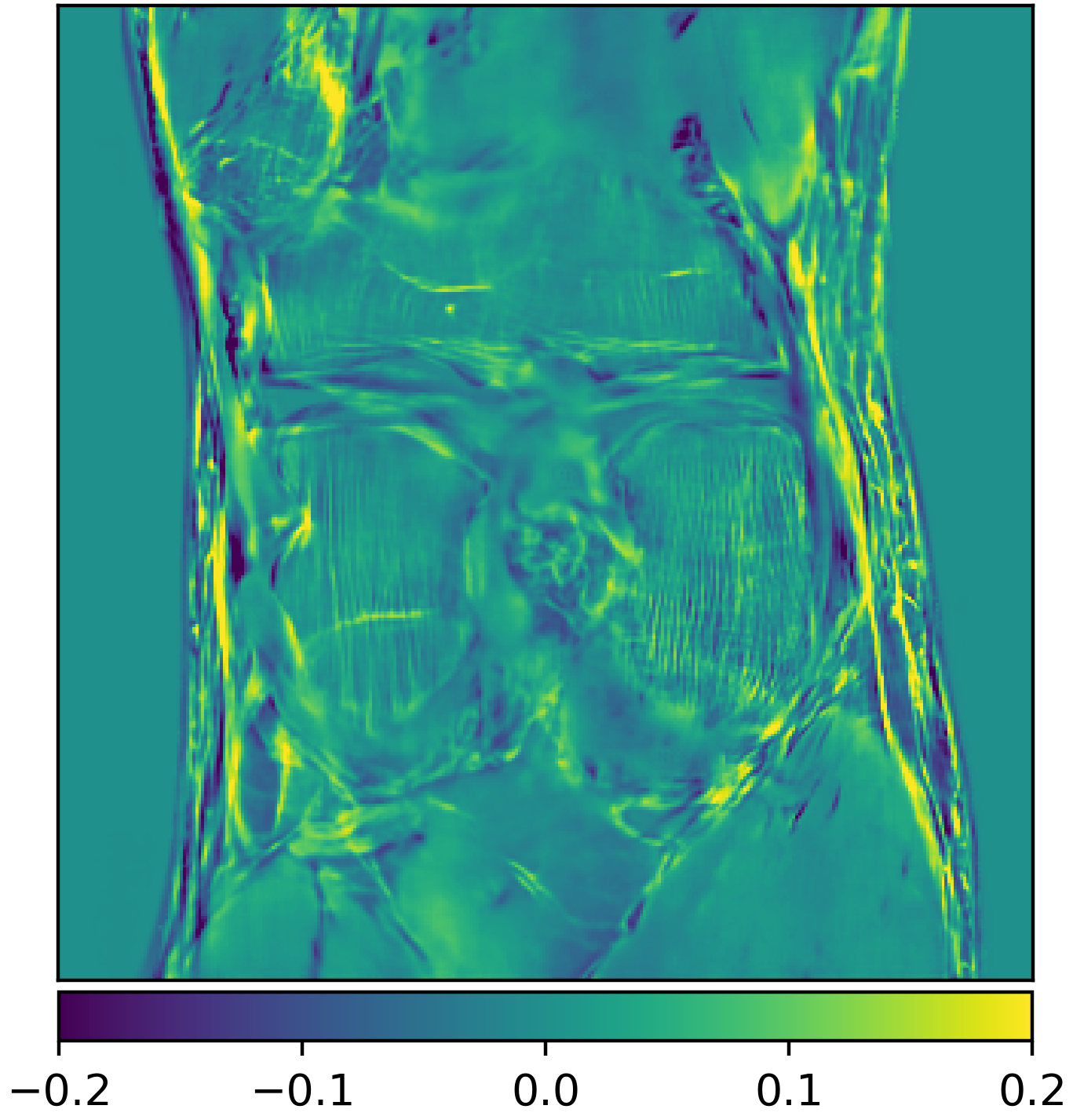}&
\includegraphics[width=0.14\textwidth]{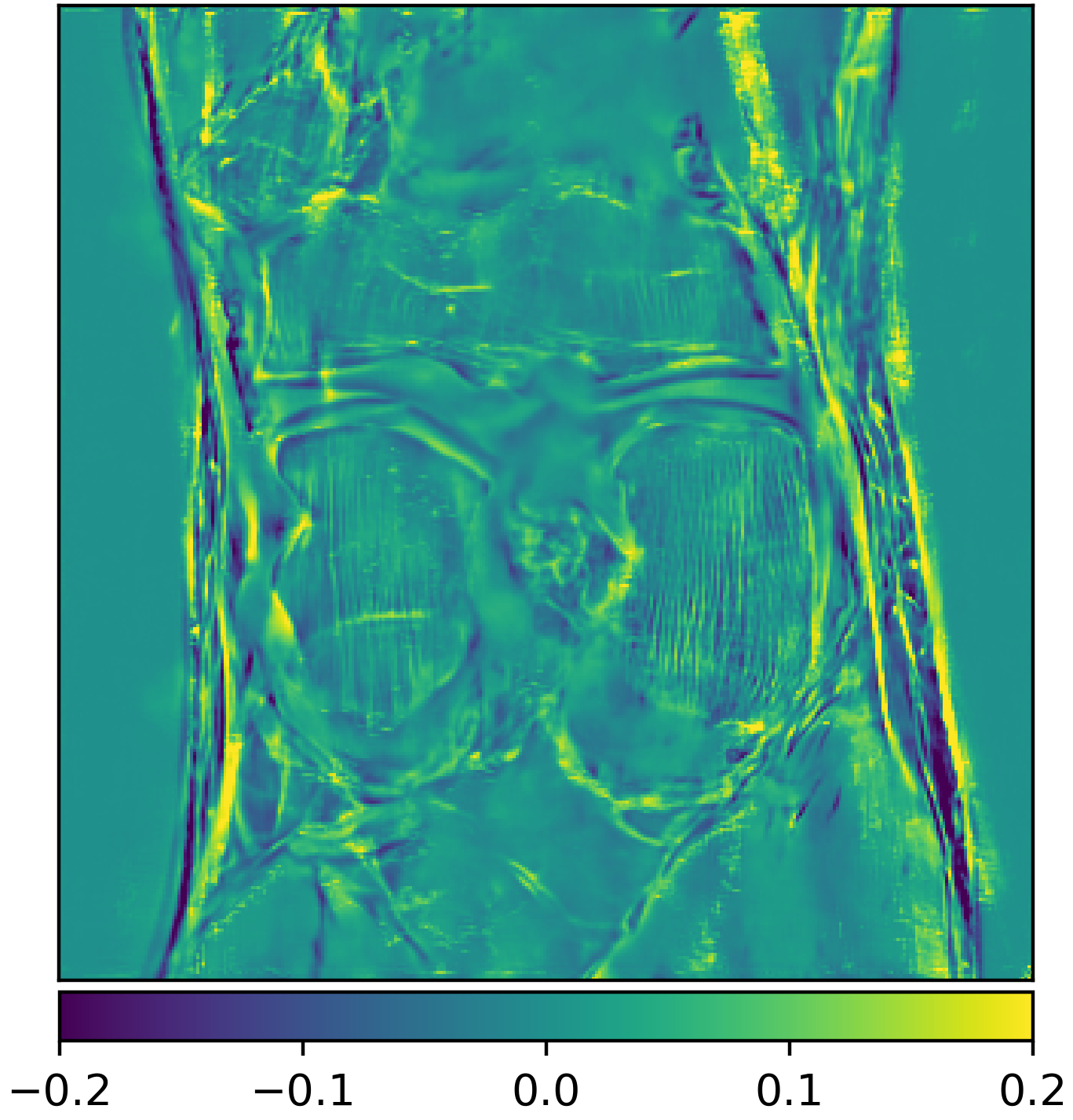}&
\includegraphics[width=0.14\textwidth]{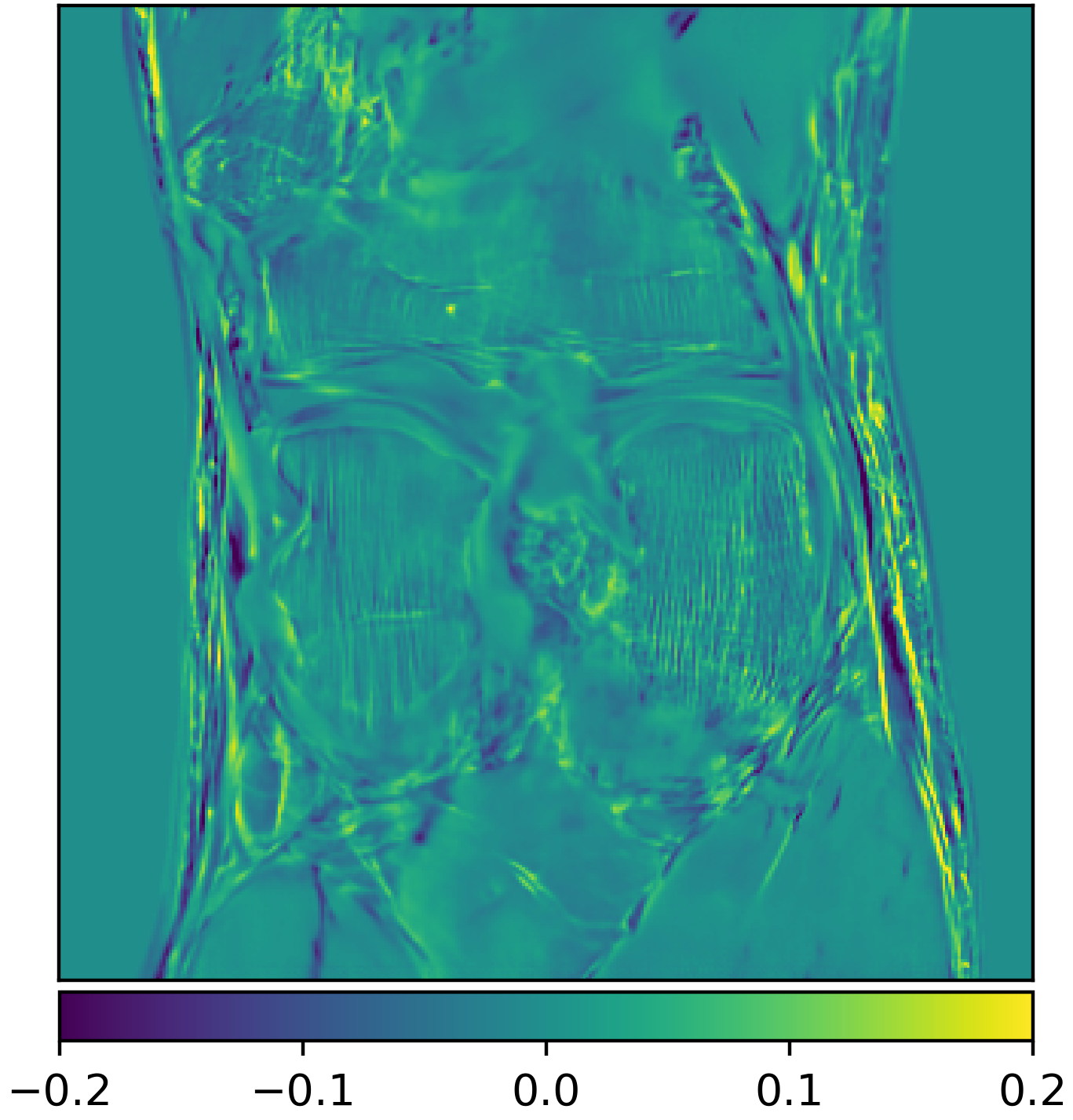}&
\includegraphics[width=0.14\textwidth]{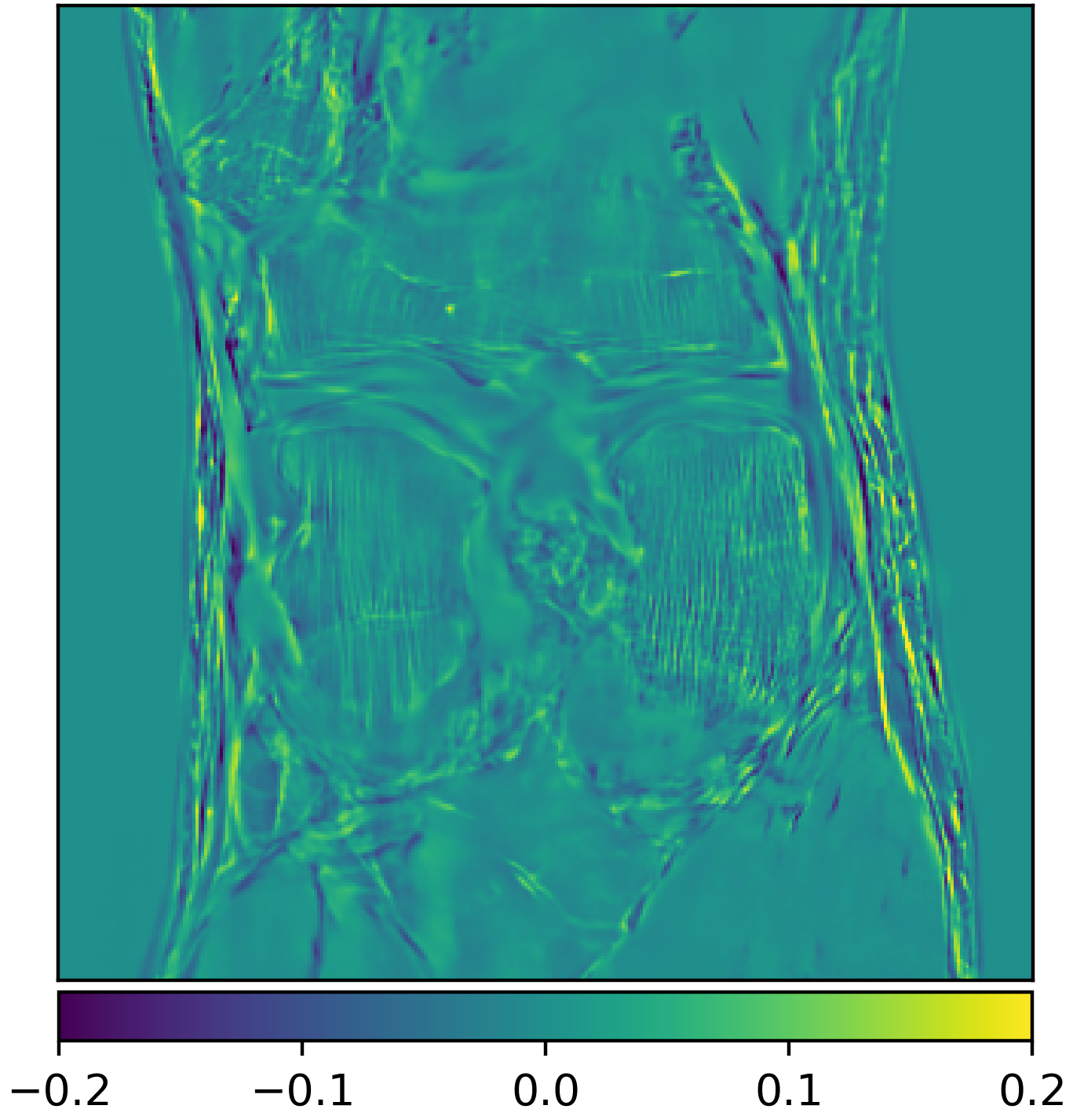} &
\includegraphics[width=0.14\textwidth]{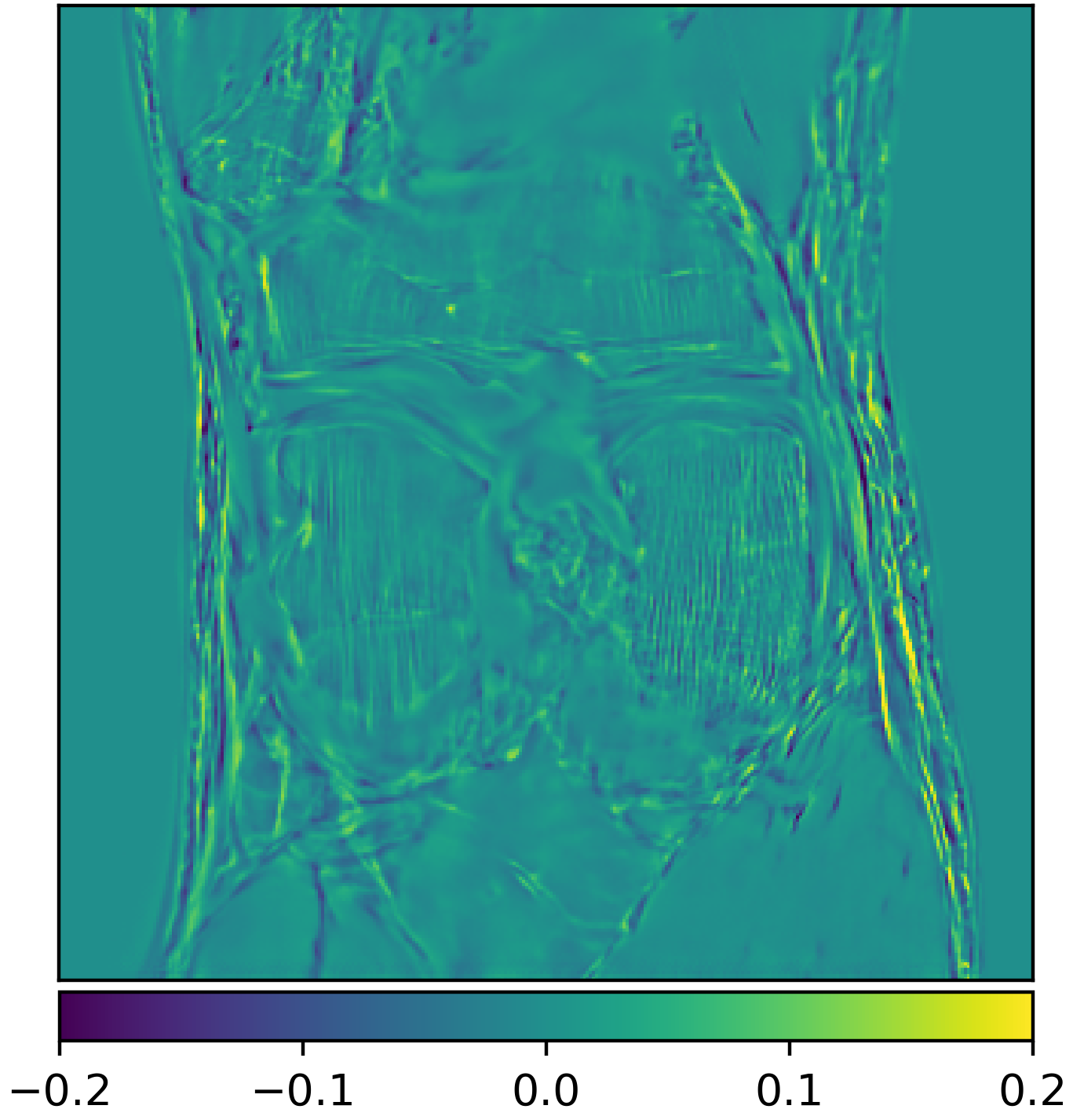} 
 \\
\small GT and Back-projected &\small AIRI: \qquad(15.55, 18.96) & \small U-Net: \qquad (14.93, 17.88) & \small NC-PDNet: \qquad (14.40, 17.65) & \small R2D2-Net (FFT): \qquad(17.46, 22.85) & \small R2D2: \qquad(18.44, 23.15) & \small R2D2-Net (NUFFT): (19.39, 25.10)\\
\end{tabular}
\vspace{-5pt}
\caption{\textbf{Reconstructed images in the first row and corresponding differences compared to the GT image in the second row by different methods for one of the validation samples.} Here, DR = 167 and AF = 16. SNR and logSNR values (dB) are displayed below.}
\vspace{-20pt}
\label{fig:vis}
\end{figure*}

\subsection{Details of the data generation}
In training, we randomly selected $N_{\text{s}}$ from the integer set $\{8,9,\dots, 79, 80\}$, corresponding to the AFs from 4 to 40, to generate k-space trajectories. We simulated k-space measurements using these radial trajectories and added corresponding noise as shown in \fref{fig:dataset} (b). For each GT image, we generated one back-projected image with a randomly selected sampling pattern to construct pairs of samples for supervised learning. In testing, we randomly selected 20 GT images from the validation dataset and then applied sampling patterns that vary the number of spokes from 10 to 80 with a step of 10, resulting in a total of 160 inverse problems.

\section{Experimental Results} \label{sec:results}
\subsection{Comparison methods}
We compare the proposed methods, including R2D2, R2D2-Net (NUFFT), and R2D2-Net (PSF), to the state-of-the-art imaging methods, from the vanilla DNN baseline U-Net \cite{ronneberger2015u}, to the unrolled DNN baseline NC-PDNet \cite{ramzi2022nc}, and to the advanced PnP algorithm AIRI\footnote{AIRI is a PnP algorithm based on a Forward-Backward optimisation structure. Its denoisers are trained to satisfy firm a nonexpansiveness constraint necessary to PnP converge \cite{pesquet2021learning}. It leverages a shelf of denoisers trained for a range noise levels, which are used adaptively across the iteration process. The combination of its convergence guarantees and its adaptive noise level functionality was shown to deliver precision and robustness superior to more basic PnP approaches. For the purpose of this comparison, the denoisers were trained on the same GT dataset as the DNNs of the R2D2 family.}, also recently introduced in astronomical imaging \cite{Matthieu2021, terris2023plug}.

\subsection{Implementation details}
The DNN models were implemented using PyTorch \cite{paszke2019pytorch}. We used the PyTorch implementation of NUFFT from \cite{muckley20}, which benefits from GPU acceleration. 
We utilized the Adam optimizer \cite{kingma2014adam} with a basic learning rate of 0.0001.  In line with the previous work \cite{aghabiglou2023ultra}, the number of output channels of the first convolution layer for U-Net is set as 64. All DNN models, including the denoisers in AIRI, were trained using a single NVIDIA Tesla V100-SXM2-16GB GPU with a batch size of 4 on Cirrus 4, a high-performance computing system. They were initialized randomly and trained until convergence. The same GPU was utilized in reconstruction for DNN models, while AIRI's iterative reconstruction process employed the GPU combined with the dual Intel 18-core Xeon E5-2695 processor on Cirrus 4. 
\begin{figure}
  \centering 
  \subfigure[SNR]{
  \begin{minipage}{.22\textwidth}
    \centering
    \includegraphics[width=\linewidth]{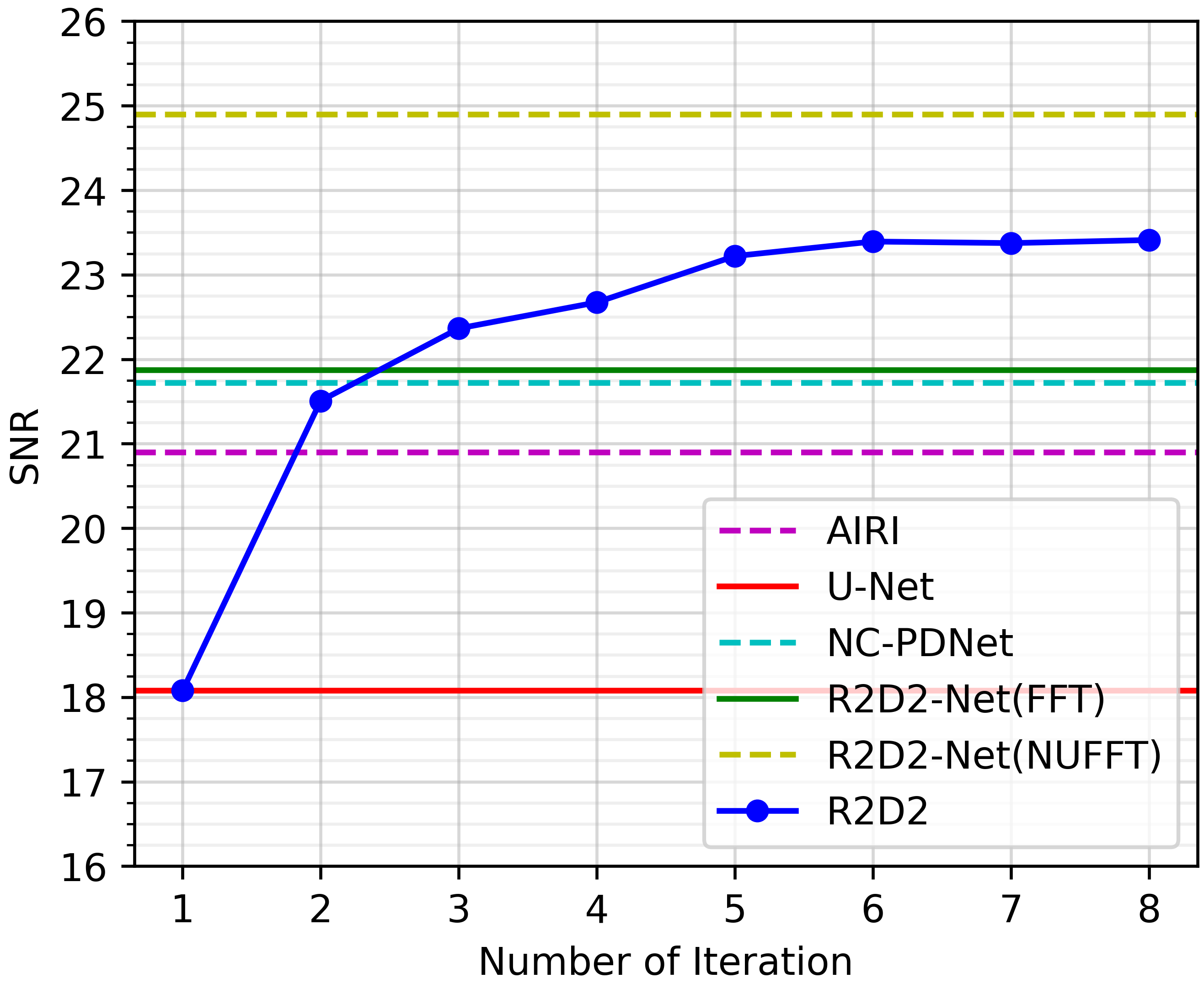}
  \end{minipage}%
  } \vspace{-5pt}
  \subfigure[logSNR]{
  \begin{minipage}{.22\textwidth}
    \centering
    \includegraphics[width=\linewidth]{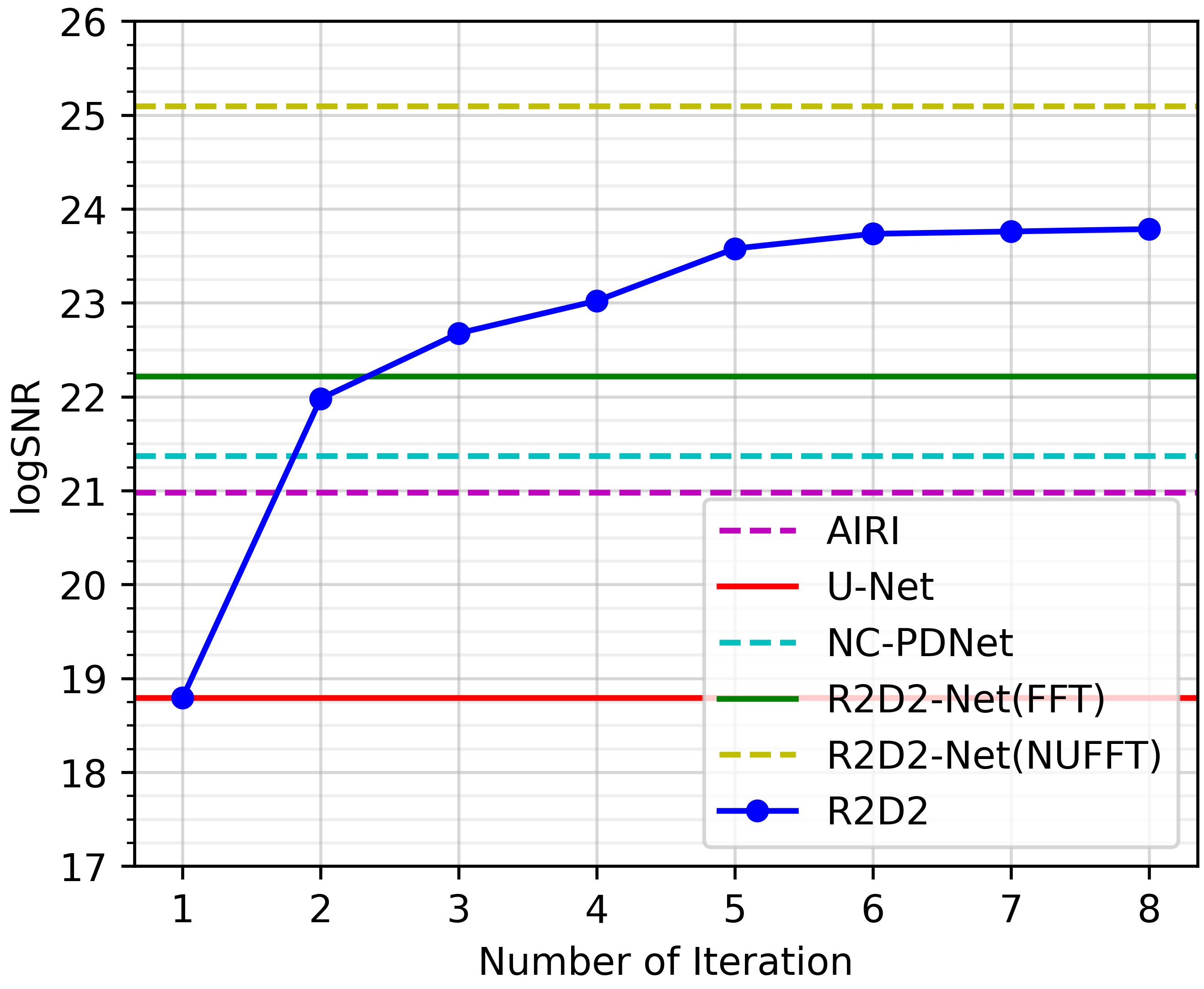}
  \end{minipage}
  } \vspace{-5pt}
  \subfigure[SNR]{
  \begin{minipage}{.22\textwidth}
    \centering
    \includegraphics[width=\linewidth]{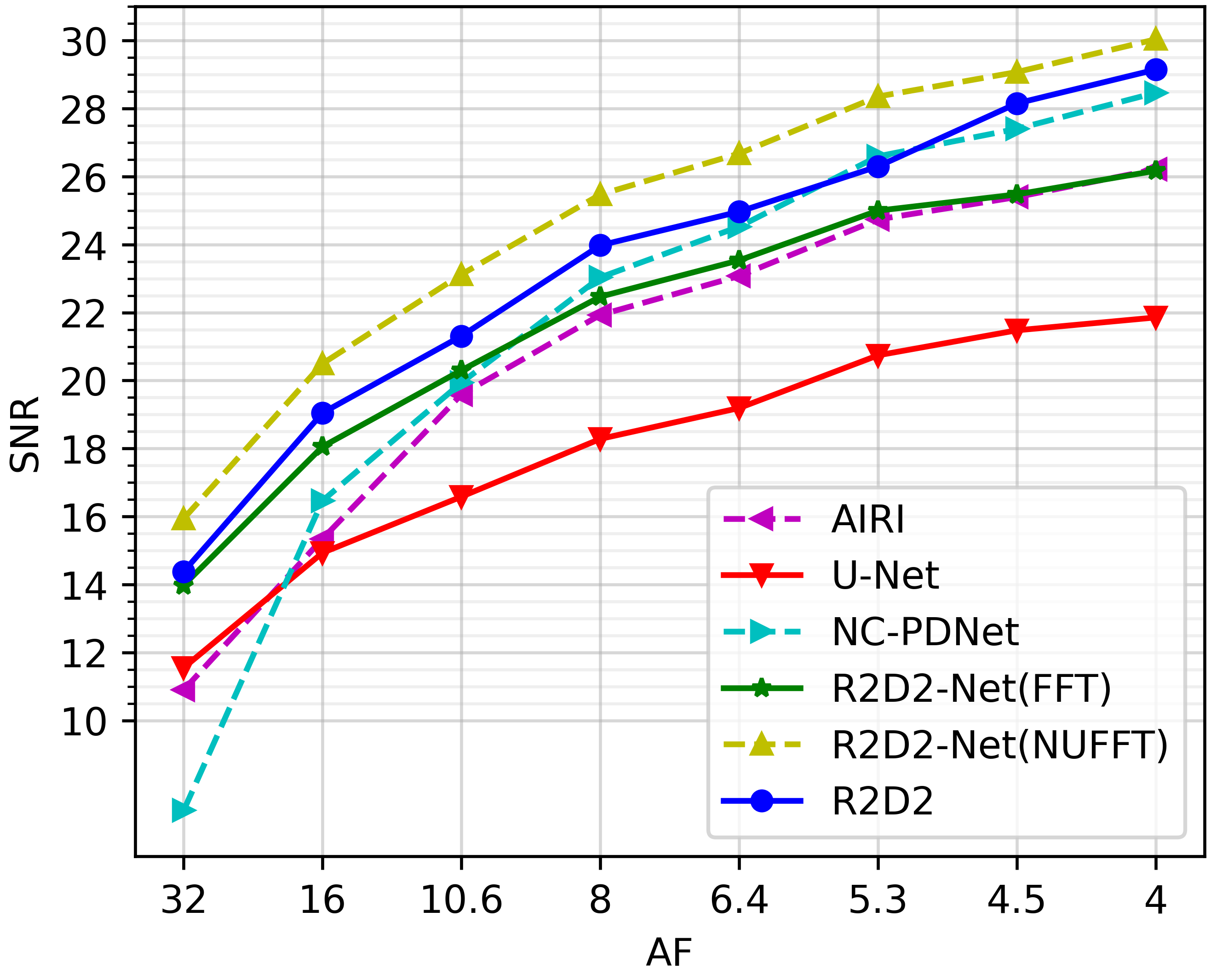}
  \end{minipage}
  }
  \hspace{-6pt} 
  \subfigure[logSNR]{
  \begin{minipage}{.22\textwidth}
    \centering
    \includegraphics[width=\linewidth]{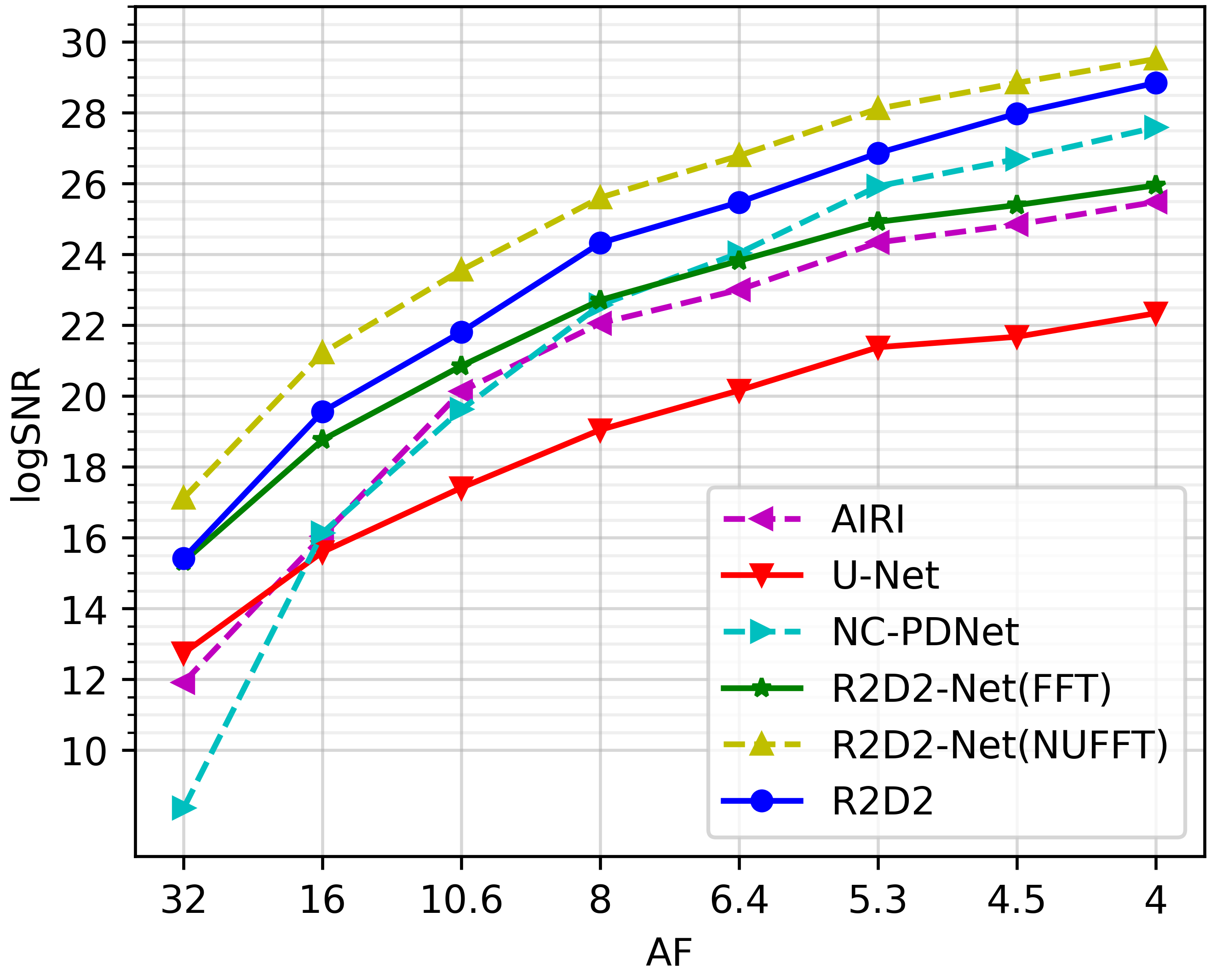}
  \end{minipage}
  }
\vspace{-5pt}
  \caption{\textbf{The reconstruction performances of considered methods across (a), (b) the number of iterations, and (c), (d) the number of spokes.} The scalable methods are denoted by continuous lines, while non-scalable methods are denoted by dotted lines.}
  \vspace{-10pt}
  \label{fig:curves}
\end{figure}

\subsection{Evaluation metrics}
We adopt linear and logarithmic signal-to-noise ratio metrics to evaluate imaging quality. Firstly, Signal-to-Noise Ratio (SNR) is defined as $\operatorname{SNR}(\bm{x}, \bm{\bar{x}})=20 
 \log_{10}\left(\|\bm{\bar{x}}\|_{2} / \|\bm{\bar{x}}-\bm{x}\|_{2} \right)$,
where $\| \cdot \|_{2}$ is the L2 norm. 
Secondly, based on the logarithmic mapping of the images parameterized by $a > 0$ is $\operatorname{rlog}\colon \bm{x} \mapsto \operatorname{log}_{a}(a\bm{x}+1)$, the logarithmic SNR (logSNR) to evaluate the ability of recovering faint signals is defined as
$
\operatorname{logSNR}(\bm{x}, \bm{\bar{x}})=\operatorname{SNR}(\operatorname{rlog}(\bm{x}), \operatorname{rlog}(\bm{\bar{x}}))
$,
where $a$ is set as the DR of the image. 

\subsection{Results}
\fref{fig:curves} (a) and (b) depict the results in terms of SNR and logSNR, respectively, averaged from the 160 inverse problems. We categorize the considered methods into two groups: scalable and non-scalable, based on their capability to be applied to large-dimensional scenarios. As the number of iterations increases, the performance exhibits a rising and converging trend for R2D2, surpassing all other algorithms aside R2D2-Net (NUFFT). \fref{fig:curves} (c) and (d) show the results across the AFs, averaged from the 20 inverse problems for each sampling pattern. NC-PDNet performs poorly in high-AF scenarios due to the simple subnetwork structure. 
As expected, R2D2-Net (NUFFT) delivers superior outcomes compared to R2D2 due to the joint training of its networks. However, its non-scalability to large dimensions renders it akin to an oracle. 
\begin{table}
\centering
\small
\caption{\small Additional acceleration ratios for R2D2.}
\begin{tabular} {@{\hspace{0.6\tabcolsep}}c|@{\hspace{0.6\tabcolsep}}c@{\hspace{0.6\tabcolsep}}|@{\hspace{0.6\tabcolsep}}c@{\hspace{0.6\tabcolsep}}|@{\hspace{0.6\tabcolsep}}c@{\hspace{0.6\tabcolsep}}|@{\hspace{0.6\tabcolsep}}c@{\hspace{0.6\tabcolsep}}}
\toprule
Algorithm$\backslash$SNR & 16  &21  &26 &Scalability\\
\midrule
U-Net &  1.92& 2.17&-&Yes\\
AIRI &  1.69 & 1.21 &1.35&No\\
R2D2-Net (FFT) &  1.15 & 1.14& 1.35&Yes\\
NC-PDNet &  1.54& 1.14& 0.98&No\\
R2D2-Net (NUFFT) &  0.77 & 0.76 & 0.79&No\\
\bottomrule
\end{tabular}
\label{tab:comparison_table_AF}
\vspace{-15pt}
\end{table}
Based on \fref{fig:curves} (c), we provide the additional acceleration ratio for R2D2 to show its capability of reducing scan time compared to other methods in \tref{tab:comparison_table_AF}. The ratio is defined as $\text{AF}_{\text{R2D2}}/\text{AF}_{\text{Algo}}$, where $\text{AF}_{\text{R2D2}}$ is the AF of R2D2 and $\text{AF}_{\text{Algo}}$ is the AF of another algorithm at the same imaging quality (SNR). In \fref{fig:vis}, R2D2 and R2D2-Net (NUFFT) provide the best reconstruction results in terms of metrics and visual performance, while only subtle differences between R2D2 and R2D2-Net (NUFFT) can be observed in the zoomed areas. 

\begin{table}
\centering
\small
\caption{\small The number of parameters, TTs, ITs, and the number of iterations for the different methods considered.}
\begin{tabular}[ht] {@{\hspace{0.1\tabcolsep}}l@{\hspace{1.1\tabcolsep}}c@{\hspace{1.1\tabcolsep}}c@{\hspace{1.1\tabcolsep}}c@{\hspace{1.1\tabcolsep}}c@{\hspace{0.1\tabcolsep}}}
\toprule
Algorithm & Par. (M) &TT (h) & IT (s)  & Iteration\\
\midrule
AIRI &  $0.6$ &$48$ & $1857 \pm 387$ & $616\pm 138$  \\
\midrule
U-Net  & $31$ &$52$ & $0.053 \pm 0.007$ & $1$\\ 
R2D2-Net (FFT)  & $248$ &$152$ & $0.129 \pm 0.004$ & $1$\\ 
R2D2 & $248$ &$140$ & $0.237\pm 0.005$ & $8$\\
\midrule
NC-PDNet  & $1.6$ & $230$ & $0.263\pm 0.011$ & $1$\\ 
R2D2-Net (NUFFT)  & $248$ & $315$ & $0.237\pm 0.005$ & $1$\\ 
\bottomrule
\end{tabular}
\label{tab:comparison_table}
\vspace{-15pt}
\end{table}

\tref{tab:comparison_table} provides the number of trainable parameters, Training Times (TTs), Inference Times (ITs) and number of iterations, averaged from the 160 inverse problems. We classify the considered methods into three categories: the PnP algorithm, scalable DNN models and non-scalable DNN models, delineated by a horizontal line in the table. The TTs of scalable DNNs are notably shorter compared to non-scalable ones. Specifically, when comparing the TT of R2D2-Net (NUFFT) to R2D2-Net (FFT), we observe an increase of approximately 163 hours, primarily due to the NUFFT-related computation during training. In terms of ITs, R2D2 is four orders of magnitude faster than AIRI, attributed to its smaller number of iterations.

\section{Conclusion}\label{sec:conclusion}
We have introduced the R2D2 deep learning-based image reconstruction paradigm to MRI, leveraging a series of end-to-end DNNs in a ``Matching Pursuit'' flavour. Each network in the series utilizes the back-projected data residual and previous estimate to enhance the reconstruction. Two unrolled variants were proposed: R2D2-Net (NUFFT), utilizing NUFFT-based data-consistency layers, and R2D2-Net (FFT), employing the FFT approximation for data consistency. For this proof of concept, and without loss of generality, we have concentrated on the single-coil scenario to facilitate a comparison between scalable and non-scalable methods. Preliminary simulations on magnitude MR images with single-coil radial sampling demonstrate that R2D2 achieves: (i) suboptimal performance compared to its unrolled counterpart R2D2-Net (NUFFT), which is non-scalable due to embedded NUFFT-based layers; (ii) superior reconstruction quality to R2D2-Net (FFT); (iii) superior reconstruction quality to state-of-the-art methods, including AIRI and NC-PDNet. 

Future research includes: (i) studying more evolved incarnations than those considered, in particular investigating advanced DNN architectures in lieu of R2D2's U-Net core module; (ii) accounting for the complex-valued nature of real-world MRI data; (iii) confirming R2D2's practical performance in large-dimensional scenarios, including multi-coil settings, and for 3D and 4D MRI. It is worth mentioning that preliminary experiments suggest that embedding the NUFFT into R2D2-Net already becomes impractical for training in a 2D multi-coil setting with 32 coils at the same image sizes as those considered here.


\bibliographystyle{IEEEbib}
\bibliography{ref}

\end{document}